%% file: ms.tex
\documentclass[conference]{IEEEtran}
\IEEEoverridecommandlockouts
\usepackage{amsmath,amssymb,amsfonts}
\usepackage{algorithmic}
\usepackage{graphicx}
\usepackage{textcomp}
\usepackage{xcolor}
\usepackage[
    backend=biber,
    natbib=true,
    url=true, 
    doi=false,
    eprint=false,
    maxcitenames=99,
    maxnames=99,
    giveninits=true
]{biblatex}

\usepackage[hidelinks,breaklinks,bookmarks=false,draft]{hyperref}
\usepackage{booktabs}
\usepackage{xcolor}
\usepackage{balance}

\usepackage{xspace}

\usepackage{amsmath} 
\usepackage{amsthm}
\usepackage{amssymb}

\usepackage[disable]{todonotes}

\usepackage[mode=buildnew]{standalone}
\usepackage{pgfplots}

\input{definitions}

\begin{document}

\title{DeXTT: Deterministic Cross-Blockchain \\ Token Transfers}

\author{\IEEEauthorblockN{Michael Borkowski\IEEEauthorrefmark{1}, Marten Sigwart\IEEEauthorrefmark{1}, Philipp Frauenthaler\IEEEauthorrefmark{1}, Taneli Hukkinen\IEEEauthorrefmark{3}, Stefan Schulte\IEEEauthorrefmark{1}}\vspace{2mm}

	\IEEEauthorblockA{
		\begin{tabular}{p{7cm} p{7cm}}
			\centering\begin{tabular}{c}
				\IEEEauthorrefmark{1}
				Distributed Systems Group\\
				TU Wien, Vienna, Austria\\
				\{m.borkowski, m.sigwart, p.frauenthaler, \\ s.schulte\}@infosys.tuwien.ac.at
			\end{tabular}
 & \centering\begin{tabular}{c}
				\IEEEauthorrefmark{3}
				Pantos GmbH \\
				Vienna, Austria \\
				contact@pantos.io \\~
			\end{tabular}
		\end{tabular}
	}
}

\maketitle

\begin{abstract}
\input{sections/00_abstract.tex}
\end{abstract}


\input{sections/10_intro.tex}
\input{sections/20_background.tex}

\input{sections/30_approach.tex}
\input{sections/40_eval.tex}

\input{sections/50_rw.tex}
\input{sections/60_conclusion.tex}

\section*{Acknowledgments}

The work presented in this paper has received funding from Pantos GmbH within the TAST research project.

\vspace{1mm}

\balance
\printbibliography

\end{document}

%% file: definitions.tex
\newcommand{\chain}{\mathcal{C}\xspace}
\newcommand{\wallet}{\mathcal{W}\xspace}
\newcommand{\cwallet}[2]{{{\chain_{#1}}:\wallet_{#2}}\xspace}

\newcommand{\xfer}[3]{\wallet_{#1} \xrightarrow{#3} \wallet_{#2}\xspace}

\newcommand{\trans}{\textsc{trans}\xspace}
\newcommand{\claim}{\textsc{claim}\xspace}
\newcommand{\contest}{\textsc{contest}\xspace}
\newcommand{\finalize}{\textsc{finalize}\xspace}
\newcommand{\veto}{\textsc{veto}\xspace}
\newcommand{\finveto}{\textsc{finalize-veto}\xspace}

\newcommand{\spend}{\textsc{spend}\xspace}

\newcommand{\chains}[4]{
{\small
\noindent
\makebox[\columnwidth]{%
\parbox{\columnwidth}{%
\setlength{\fboxsep}{5pt}%
\fbox{%
\begin{minipage}[t][#1]{.28\columnwidth}
\centerline{\underline{Blockchain $\chain_a$}} \vspace*{2mm}
#2
\end{minipage}}%
\hfill
\fbox{%
\begin{minipage}[t][#1]{.28\columnwidth}
\centerline{\underline{Blockchain $\chain_b$}} \vspace*{2mm}
#3
\end{minipage}}%
\hfill
\fbox{%
\begin{minipage}[t][#1]{.28\columnwidth}
\centerline{\underline{Blockchain $\chain_c$}} \vspace*{2mm}
#4
\end{minipage}}}}
\vspace*{1mm}
}
}

\newcommand{\va}[1]{#1}
\newcommand{\vaa}[1]{\phantom{1}#1}

\newcommand{\bbb}[3]{%
	\begin{tabbing}%
	\hspace*{2.1cm}\=\kill%
	~~$\wallet_s$ balance: \> #1 \\%
	~~$\wallet_d$ balance: \> #2 \\%
	~~$\wallet_w$ balance: \> #3 \\[3mm]%
	\end{tabbing}\vspace*{-6mm}%
}

%% file: sections/00_abstract.tex
Current blockchain technologies provide very limited interoperability. Restrictions with regards to asset transfers and data exchange between different blockchains reduce usability and comfort for users, and hinder novel developments within the blockchain space. 

As a first step towards cross-blockchain interoperability, we propose the DeXTT cross-blockchain transfer protocol, which can be used to transfer a token on any number of blockchains simultaneously in a decentralized manner. We provide a reference implementation using Solidity, and evaluate its performance. We show logarithmic scalability of DeXTT with respect to the number of participating nodes, and analyze cost requirements of the transferred tokens.

%% file: sections/10_intro.tex
\section{Introduction}
\label{sec:intro}

Blockchains, the underlying technology of cryptocurrencies, have gained significant interest in both industry and research~\cite{zohar2015bitcoin}. After the feasibility of decentralized ledgers has been demonstrated by Bitcoin~\cite{nakamoto2008bitcoin}, significant investment into research and development related to blockchains and cryptocurrencies was sparked. Technologies range from adding new layers on top of existing blockchain implementations~\cite{counterparty,omnilayer}, improvements of Bitcoin itself~\cite{ltc}, to entirely new blockchains~\cite{ethereum-yellowpaper}, which provide novel concepts, such as smart contracts~\cite{7467408}.

The level of investment in the blockchain space is indicative for the technological impact and the broad range of potential use cases for blockchain technologies~\cite{fernandez2018review}. However, despite this positive momentum, structural problems exist within the blockchain field. Development so far is centered on the creation of new blockchains and currencies, or altering major blockchains like Bitcoin~\cite{yli2016current}. Furthermore, there is substantial research on potential use cases of blockchains in various economic, social, political, and engineering fields~\cite{fernandez2018review}. Nevertheless, the ways in which blockchains could potentially interact with each other remain mostly unexplored.

The constant increase in the number of independent, unconnected blockchain technologies causes significant fragmentation of the research and development field, and poses challenges for both users and developers of blockchain technologies. On the one hand, users have to choose which currency and which blockchain to use. Choosing novel, innovative blockchains enables users to utilize new features and take advantage of state-of-the-art technology. However, users also risk the loss of funds if the security of such a novel blockchain is subsequently breached, potentially leading to a total loss of funds~\cite{nofer2017blockchain}. Choosing mature, well-known blockchains reduces the risk of such loss, since these blockchains are more likely to have been analyzed in-depth~\cite{li2017survey}, but novel features remain unavailable.

On the other hand, when designing decentralized block\-chain-based applications, currently, developers must decide which blockchain to base their application on. This can form a substantial impedance to research and technical progress, since individual technologies form isolated solutions, and interoperability between blockchains is mostly not given.

We therefore aim to enable cross-blockchain interoperability. As an overarching goal, we seek to provide means of interaction between blockchains, including cross-blockchain data transmission, cross-blockchain smart contract interaction, or cross-blockchain currency transfer. As a first step to enable such cross-blockchain interoperability, we propose a protocol for cross-blockchain token transfers, where the transferred token is not locked within an individual blockchain. Instead, it can be used on any number of blockchains, and its transactions are autonomously synchronized across blockchains by the system in a decentralized manner. Our solution prevents double spending, is resilient to the cross-blockchain proof problem~(XPP)~\cite{tast-wp2}, and does not need external oracles or other means of cross-blockchain communication to function. We provide a reference implementation using Solidity, and evaluate its performance with regards to time and cost.

The contributions of this manuscript are as follows:
\begin{itemize}
\item We discuss how to use eventual consistency for cross-blockchain token transfers by utilizing concepts such as claim-first transactions and deterministic witnesses.
\item We formally define \emph{Deterministic Cross-Blockchain Token Transfers}~(DeXTT), a protocol which implements eventual consistency for cross-blockchain token transfers.
\item We provide a reference implementation in Solidity, presenting and evaluating DeXTT.
\end{itemize}

The remainder of this paper is structured as follows. In Section~\ref{sec:background}, we discuss underlying technologies, provide a brief discussion of blockchains and transaction types, claim-first transactions, and witness rewards, and define notation used throughout this work. Section~\ref{sec:approach} presents the transfer protocol in detail, and Section~\ref{sec:eval} provides an evaluation of our approach. Section~\ref{sec:rw} gives a brief overview of related work, and describes the relation of the work at hand to our own former work in the field of cross-blockchain interoperability. Finally, Section~\ref{sec:conclusion} concludes the paper.

%% file: sections/20_background.tex
\section{Background}
\label{sec:background}

Our work aims at providing a protocol for cross-blockchain asset transfers, ensuring that such transfers are performed in a decentralized and trustworthy manner. Assets can be represented on blockchains in various ways. Apart from native currencies~(e.g., Ether on the Ethereum blockchain, or Bitcoin on the Bitcoin blockchain), there are other types of assets, commonly called \emph{tokens}. In the recent past, various asset types with different properties have been discussed, such as fungibility, divisibility, and types of implementation like the user-issued asset (UIA) and \emph{Unspent Transaction Output}~(UTXO) models. We refer to our previous work for a thorough analysis~\cite{tast-wp1}.

In the work at hand, we discuss a token that exists on a given number of blockchains simultaneously, i.e., a \emph{pan-blockchain token}~(denoted as \textit{PBT}). PBT are not locked to a single blockchain and can be traded using the DeXTT protocol, which ensured synchronization of token balances across blockchains. We refer to the set of blockchains participating in this protocol as an \emph{ecosystem of blockchains}. According to our protocol, a wallet $\wallet_w$ is holding PBT not only on a given blockchain, but on all blockchains in the ecosystem. Thus, a transfer from $\wallet_w$ to another wallet $\wallet_v$ is required to be recorded on all participating blockchains, and there must be consensus among all participating blockchains about the balance of each wallet.

Due to the XPP~\cite{tast-wp2}, strict consistency between blockchains is not possible using practical means, since any verification of data between two blockchains would essentially require the nodes of one blockchain to verify blocks of another blockchain. This requires both the data and the consensus protocol to be shared across blockchains, which is not possible in practice. Therefore, in our proposal, we relax this requirement to eventual consistency, i.e., we accept temporary disagreement with regard to balances, as we show in the following. In practice, blockchains themselves only provide eventual consistency, since there is no guarantee when data submitted to the network will be included in a block. Therefore, using eventual consistency for synchronizing data between blockchains is a feasible approach.

For the purpose of this paper, we follow the assumption that each party is generally interested in all the blockchains in an ecosystem, and specifically, in the consistency of their balance across all blockchains. This means that all interested parties (i.e., wallet holders) are monitoring all blockchains in the ecosystem, and if a party participates in the protocol on one blockchain, it also participates on all other blockchains. We support this assumption by defining later that any inconsistency in wallet balances between blockchains effectively renders the wallet useless.

DeXTT assumes that non-zero token balances already exist on the involved blockchains. We explicitly do not define the economic aspect of PBT, i.e., the lifecycle of tokens. Several minting strategies exist, and we provide an overview of such approaches (constant supply, minting rate, etc.) in previous work~\cite{tast-wp1}. Any of these approaches is usable together with DeXTT, since the protocol assumes that tokens already exist.

\subsection{Cross-Blockchain Balance Consistency}
\label{sec:consistency}

As outlined before, we require eventual consistency between blockchains participating in the proposed protocol. Since due to the XPP, we cannot directly propagate information across blockchains, we require an alternative way to reach consistency across blockchains.

For this, we propose to achieve eventual consistency using \emph{claim-first transactions}~\cite{tast-wp2}. While traditionally, blockchain transfers disallow claiming tokens before they have been marked as spent, we explicitly decouple the required temporal order of $\spend \rightarrow \claim$ and allow its reversal, i.e., claiming tokens before spending them. In our case, for a certain period of time, tokens are allowed to exist in the balance of both the sender and the receiver (on different blockchains), namely until the information is propagated to all blockchains. In the presented protocol, we provide a mechanism to enforce eventual spending of the tokens in the sender balance, as described in Section~\ref{sec:approach}.

In order to ensure such eventual consistency, we rely on parties observing a transfer to propagate this information across blockchains. These parties are denominated as \emph{witnesses}. A monetary incentive is provided for any witness in order to ensure propagation. We use part of the transferred PBT for these witness rewards. The main challenge of this approach is the decision which witness receives the reward. Using a first-come-first-serve basis is not feasible, since it is possible that on one blockchain, one witness is the first to propagate the transfer and claim the reward, while on another blockchain, another witness takes this place. This would lead to two different witnesses receiving a reward on two different blockchains, and therefore, to potentially inconsistent balances.

In this work, we address this problem by using \emph{deterministic witnesses}~\cite{tast-wp3}. In short, instead of using a first-come-first-serve reward distribution, we define a \emph{witness contest}. Its duration is fixed to a validity period, \emph{contestants} (i.e., reward candidates) can register for the contest, and the decision of who wins the contest is made deterministically and predictably by each blockchain at the end of the contest. In Section~\ref{sec:approach}, we propose an approach for deciding the winning witness in a way that is fair~(i.e., all contestants have the same chance of winning), while at the same time, it is purely deterministic, and---given the assumptions discussed above---assures all blockchains reach the same decision about assigning witness rewards.

Our approach therefore solves the problem of assigning witness rewards, which is required as an incentive for observers of a cross-blockchain transfer to propagate this transfer information, ensuring eventual consistency across the ecosystem of blockchains.

\subsection{Cryptographic Signatures and Hashes}
\label{sec:hashes}

In our approach, we make extensive use of cryptographic signatures and hashes, which are essential for blockchains themselves. For instance, the ECDSA algorithm~\cite{johnson2001elliptic} is used by Ethereum for creating and verifying signatures, and is also implemented natively and available to the Ethereum Virtual Machine~(EVM)~\cite{hirai2017defining}. We use Solidity, the smart contract language of Ethereum, for the reference implementation of DeXTT. However, we note that DeXTT is not limited to Solidity or the EVM, and other blockchains offering signatures and hash algorithms can very well be used. The only crucial property required by our approach is a distribution of hash values which is approximately uniform. \textsc{Keccak256}, the hash algorithm used by Ethereum, satisfies this requirement~\cite{gholipour2011pseudorandom}, as does the \textsc{Sha-256} algorithm used by Bitcoin~\cite{gilbert2003security}.

\subsection{Notations and Conventions}

In the following, we use particular notations for concise description of certain objects: We denote blockchains as $\chain$ with a subscript letter, e.g., $\chain_a$. Additionally, we denote wallets as $\wallet$ with a subscript letter, e.g., $\wallet_s$, $\wallet_d$, or $\wallet_w$. A wallet consists of a pair of corresponding keys, out of which one is a public key, and one is a private key. When referring to a token transfer in general, $\wallet_s$ is used to denote the source (sending) wallet, $\wallet_d$ is used to denote the destination (receiving) wallet, and $\wallet_w$ denotes a witness as discussed in Section~\ref{sec:consistency}. As discussed in Section~\ref{sec:intro} and demonstrated in Table~\ref{tab:initial}, the balance of a wallet is stored across all blockchains.

In this work, we use the concept of \emph{transactions} to denote actions executed on a blockchain which modify the blockchain state. We use the expression ``$\wallet_w$ posts the transaction \trans on $\chain_c$'' to describe the conceptual protocol. In a scenario where smart contracts are used, this translates to the key pair of $\wallet_w$ being used to sign a call to the smart contract on blockchain $\chain_c$, where the function \texttt{trans()} is invoked. For certain transactions, we define preconditions (e.g., sufficient balances), which can be implemented as checks within the smart contract function. The transactions posted by wallets can either originate from the action of a user, or be initiated by a program (e.g., a wallet application) acting autonomously.

To denote our transactions, we use the notation as shown in~(\ref{eq:trans}), where \trans is the transaction type used~(one out of \claim, \contest, \finalize, \veto, and \finveto), $\wallet_w$ is the wallet~(i.e., the pair of keys) used to sign and post the transaction, $a$, $b$, and $c$ denote data contained in the transaction (i.e., the arguments), and $\sigma$ is the signature when using the private key of $\wallet_w$ to sign the data $[a, b, c]$. For brevity, we use only $\sigma$ to denote a multivariate value, e.g., a three-variate ECDSA signature.

\begin{gather} \label{eq:trans}
	\begin{aligned}
	\wallet_w:\trans~&\Big[~a,\,b,\,c~\Big]_{\sigma}
	\end{aligned}
\end{gather}

We denote a transfer of $x$~PBT from $\wallet_s$ to $\wallet_d$ as \mbox{$\xfer{s}{d}{x}$}. Furthermore, we denote the PBT balance of $\wallet_w$ recorded on $\chain_c$ as $\cwallet{c}{w}$.

%% file: sections/30_approach.tex
\section{Decentralized Cross-Blockchain Transfers}
\label{sec:approach}

In the following, we present the DeXTT protocol, together with an example transaction. In our example, we consider three blockchains participating in cross-blockchain transfers, $\chain_a$, $\chain_b$, and $\chain_c$. Note, however, that our approach is applicable to an arbitrary number of blockchains. Furthermore, we consider the wallets $\wallet_s$, $\wallet_d$, $\wallet_u$, $\wallet_v$, and $\wallet_w$. We assume that initially, $\wallet_s$ has 80~PBT, and all other wallets have a balance of zero~(see Table~\ref{tab:initial}). We furthermore use a fixed reward of 1~PBT for the witness propagating this transaction across the blockchain ecosystem. Note that pro rata fees (e.g., 1\% of the transferred PBT, or an amount selected by the sender) are also possible and the exact fee model is an economic choice. We will discuss this in more detail in Section~\ref{sub:costanalysis}.

As discussed in Section~\ref{sec:consistency}, claim-first transactions require all blockchains within the ecosystem to maintain and synchronize token balances. Therefore, the initial situation is as depicted in Table~\ref{tab:initial}. Balances for $\wallet_u$ and $\wallet_v$ are not shown, as they will remain zero throughout the example.

\begin{table}
\caption{Initial State of the Involved Blockchains at $t = 0$}
\label{tab:initial}
\chains{16mm}{
	\bbb{\va{80}}{\vaa{0}}{\vaa{0}}%
}{
	\bbb{\va{80}}{\vaa{0}}{\vaa{0}}%
}{
	\bbb{\va{80}}{\vaa{0}}{\vaa{0}}%
}
\end{table}

\subsection{Transfer Initiation}

In the following, we assume that $\wallet_s$ intends to transfer $20$~PBT to $\wallet_d$, i.e., reduce the PBT balance of $\wallet_s$ by $20$, increase the PBT balance of $\wallet_d$ by $19$ ($20$ reduced by $1$, the witness reward), and increase the PBT balance of a (yet to be decided) witness wallet by $1$. As stated in Section~\ref{sec:consistency}, we only require eventual consistency for this transfer, i.e., a temporary overlap is allowed where $\wallet_d$ has already received $19$~PBT, but the balance of $\wallet_s$ is still unchanged.

Therefore, $\wallet_s$ signs this intent, confirming that indeed, $20$~PBT---minus $1$~PBT of witness reward---are to be transferred to $\wallet_d$. Furthermore, we define a validity period for the transfer, which denotes the time during which the witness selection for the transfer has to take place. In our example scenario, this time span lasts for 1 minute, however, this time can be set significantly shorter or longer, depending on the use case. We provide an analysis of the impact of this parameter in Section~\ref{sec:scalability}.

We denote the entirety of the sender's intent using the notation shown in~(\ref{eq:proof1}), where $[t_0,t_1]$ is the validity period, and $\alpha$ denotes the signature of the entire content of the brackets by $\wallet_s$. The resulting signature itself is denoted as $\alpha$. We use the ECDSA algorithm, natively supported by the EVM, for all signatures. However, other algorithms can also be used, assuming that their verification is supported on all involved blockchains.

\begin{equation}
	\Big[~\xfer{s}{d}{x},\,t_0,\,t_1~\Big]_{\alpha}
\label{eq:proof1}
\end{equation}

The data contained in~(\ref{eq:proof1}) is transferred to the receiving wallet $\wallet_d$. This transfer can happen on any blockchain within the ecosystem, or using an off-chain channel. Since all of the data contained in~(\ref{eq:proof1}) will be published throughout the DeXTT transaction, this channel does not need to be secure, and we do not specifically define any communication means. The receiving wallet then counter-signs the data from~(\ref{eq:proof1}) using its respective private key, yielding the entire \textit{Proof of Intent}~(\textit{PoI}), as shown in~(\ref{eq:poi}).

\begin{equation}
	\Big[~\xfer{s}{d}{x},\,t_0,\,t_1,\,\alpha~\Big]_{\beta}
\label{eq:poi}
\end{equation}

The PoI contains all information necessary to prove to any blockchain (i.e., to its smart contracts and miners) that the transfer is authorized by the sender and accepted by the receiver. The receiver can now post this PoI using a transaction we call \claim. This transaction allows the receiver to publish the PoI in order to later claim the transferred PBT. The receiver can post this on any blockchain within the ecosystem, and does not need to post it on more than one blockchain. The \claim transaction is defined and noted as shown in~(\ref{eq:claim-g}).

\begin{gather} \label{eq:claim-g}
	\begin{aligned}
	\wallet_d:\claim~&\Big[~\xfer{s}{d}{x},\,t_0,\,t_1,\,\alpha~\Big]_{\beta}
	\end{aligned}
\end{gather}

The preconditions for the \claim transaction are (i)~that the PoI is valid (i.e., that the signatures $\alpha$ and $\beta$ are correct), (ii)~that the balance of the source wallet $\wallet_s$ is sufficient, (iii)~that the PoI is not expired, i.e., that its $t_1$ has not yet passed ($t < t_1$), and (iv)~that no PoI is known to the blockchain on which it is posted with an overlapping validity period and the same source wallet $\wallet_s$. In other words, a wallet must not sign an outgoing PoI while another outgoing PoI is still pending. This is done in order to prevent a double spending attack, where two PoIs are signed which are conflicting, i.e., which, if both were executed, would reduce the sender's balance below zero.

The purpose of the \claim transaction is the publishing of the PoI, which can then be propagated across the blockchain ecosystem as described later.

In our example, we assume that the receiver $\wallet_d$ posts the \claim transaction (containing the PoI) on $\chain_a$ as shown in~(\ref{eq:claim-x}), where 1 and 61 mark the validity period in seconds (i.e., one minute total validity), \texttt{0xAA} is assumed to be the signature~$\alpha$, and \texttt{0xBB} is assumed to be the signature $\beta$. For brevity, one-byte signatures are used for demonstration in this example. Naturally, in reality, the signature hashes are longer (e.g., 32 bytes for \textsc{Keccak256}).

\begin{gather} \label{eq:claim-x}
	\begin{aligned}
	\wallet_d:\claim~&\Big[~\xfer{s}{d}{20},\,1,\,61,\,\texttt{0xAA}~\Big]_{\texttt{0xBB}}
	\end{aligned}
\end{gather}

The \claim transaction on $\chain_a$ changes the blockchain state as shown in Table~\ref{tab:claim}. We see that the PoI has been stored within $\chain_a$, which is referred to by its signature $\alpha$. The balances remain unchanged on $\chain_a$ because the validity period is not yet concluded, i.e., $t_1$ is not yet reached. Naturally, since no information has been posted yet to $\chain_b$ and $\chain_c$, these blockchains also remain unchanged at this point.

\begin{table}
\caption{State after PoI Publication at $t = 1$}
\label{tab:claim}
\chains{30mm}{
	\bbb{\va{80}}{\vaa{0}}{\vaa{0}} %
	PoI \texttt{0xAA}: \\[1mm]
	\hspace*{2mm} $\xfer{s}{d}{20}$ \\
	\hspace*{2mm} $t_1 = 61$
}{
	\bbb{\va{80}}{\vaa{0}}{\vaa{0}}%
}{
	\bbb{\va{80}}{\vaa{0}}{\vaa{0}}%
}
\end{table}

\subsection{Witness Contest}

At this point, the information about the intended transfer~(the PoI) is only recorded on $\chain_a$. However, this information must be propagated to all other blockchains as well to ensure consistency of balances across blockchains. We use the following mechanism, which we refer to as the \emph{witness contest}, to ensure this consistency.

Any party observing the \claim transaction on $\chain_a$ can become a contestant, i.e., a candidate for receiving a reward. In order to become a contestant, the party must propagate the PoI across all blockchains in the ecosystem. We define the transaction used for this as \contest. This transaction is defined for any arbitrary wallet $\wallet_o$ as shown in~(\ref{eq:contest}), where the new signature $\omega$ is the result of the contestant $\wallet_o$ signing the PoI. This signature will later play a role in determining the winner of the witness contest, as described in Section~\ref{sec:deterministic-witness-selection}.

\begin{gather} \label{eq:contest}
	\begin{aligned}
	\wallet_o:\contest~&\Big[~\xfer{s}{d}{x},\,t_0,\,t_1,\,\alpha,\,\beta~\Big]_{\omega}
	\end{aligned}
\end{gather}

The \contest transaction can be posted multiple times by various contestants during the validity period. The preconditions are the same as for the \claim transaction, i.e., the PoI must be valid and must not violate any other PoI's validity period. The only effect of the \claim transaction is that the PoI itself and the contestant's participation in the witness contest are recorded on the respective blockchain.

In our example, we assume that $\wallet_u$ is the first to post a \contest transaction on $\chain_b$ as shown in~(\ref{eq:contest-1}), where again, $1$ and $61$ denote the validity period, \texttt{0xAA} and \texttt{0xBB} are the PoI signatures, and \texttt{0xC2} is the signature resulting from $\wallet_u$ signing the PoI. The signature values in this example are chosen arbitrarily in order to demonstrate the subsequent witness contest. Again, one-byte signatures are used for brevity.

\begin{gather} \label{eq:contest-1}
	\begin{aligned}
	\wallet_u:\contest~&\Big[~\xfer{s}{d}{20},\,1,\,61,\,\texttt{0xAA},\,\texttt{0xBB}~\Big]_{\texttt{0xC2}}
	\end{aligned}
\end{gather}

Next, we assume that the other observers $\wallet_v$ and $\wallet_w$ become contestants by posting similar \contest transactions. We assume that the resulting signature $\omega$ for $\wallet_v$ is \texttt{0xC3}, and that the signature for $\wallet_w$ is \texttt{0xC1}.

\begin{gather} \label{eq:contest-2}
	\begin{aligned}
	\wallet_v:\contest~&\Big[~\xfer{s}{d}{20},\,1,\,61,\,\texttt{0xAA},\,\texttt{0xBB}~\Big]_{\texttt{0xC3}}
	\end{aligned}
\end{gather}
\begin{gather} \label{eq:contest-3}
	\begin{aligned}
	\wallet_w:\contest~&\Big[~\xfer{s}{d}{20},\,1,\,61,\,\texttt{0xAA},\,\texttt{0xBB}~\Big]_{\texttt{0xC1}}
	\end{aligned}
\end{gather}

Transactions (\ref{eq:contest-1}--\ref{eq:contest-3}) are eventually posted to $\chain_a$, $\chain_b$, and $\chain_c$. This is because every contestant participating in the contest is interested in participating in all blockchains in the ecosystem to maintain their own consistency.

The state resulting from the three contestants posting to $\chain_a$, $\chain_b$, and $\chain_c$ is shown in Table~\ref{tab:contest}. The blockchain maintains a list of contestants together with their $\omega$ signature values.

\begin{table}
\caption{State During Witness Contest at $t = 2$}
\label{tab:contest}
\chains{44mm}{
	\bbb{\va{80}}{\vaa{0}}{\vaa{0}} %
	PoI \texttt{0xAA}: \\[1mm]
	\hspace*{2mm} $\xfer{s}{d}{20}$ \\
	\hspace*{2mm} $t_1 = 61$ \\
	\hspace*{2mm} Contestants: \\
	\hspace*{2mm} $\wallet_u$ (\texttt{0xC2}) \\
	\hspace*{2mm} $\wallet_v$ (\texttt{0xC3}) \\
	\hspace*{2mm} $\wallet_w$ (\texttt{0xC1})
}{
	\bbb{\va{80}}{\vaa{0}}{\vaa{0}} %
	PoI \texttt{0xAA}: \\[1mm]
	\hspace*{2mm} $\xfer{s}{d}{20}$ \\
	\hspace*{2mm} $t_1 = 61$ \\
	\hspace*{2mm} Contestants: \\
	\hspace*{2mm} $\wallet_u$ (\texttt{0xC2}) \\
	\hspace*{2mm} $\wallet_v$ (\texttt{0xC3}) \\
	\hspace*{2mm} $\wallet_w$ (\texttt{0xC1})
}{
	\bbb{\va{80}}{\vaa{0}}{\vaa{0}} %
	PoI \texttt{0xAA}: \\[1mm]
	\hspace*{2mm} $\xfer{s}{d}{20}$ \\
	\hspace*{2mm} $t_1 = 61$ \\
	\hspace*{2mm} Contestants: \\
	\hspace*{2mm} $\wallet_u$ (\texttt{0xC2}) \\
	\hspace*{2mm} $\wallet_v$ (\texttt{0xC3}) \\
	\hspace*{2mm} $\wallet_w$ (\texttt{0xC1})
}
\end{table}

\subsection{Deterministic Witness Selection}
\label{sec:deterministic-witness-selection}

After the expiration of $t_1$, the witness contest ends, a winning witness must be selected, and awarded the witness reward. This is performed by the \finalize transaction, which must be triggered after $t_1$.

Conceptually, this transaction is purely time-based. It can be triggered by the receiver, by any other party, or using a decentralized solution like the \emph{Ethereum Alarm Clock}~\cite{berg2018chronos}. The latter approach has the advantage of being independent of any party's activity. However, for simplicity, in our current approach and the discussion below, we assume that the destination wallet $\wallet_d$ posts the \finalize transaction on each blockchain. The \finalize transaction is defined as shown in~(\ref{eq:finalize}).

\begin{gather} \label{eq:finalize}
	\begin{aligned}
	\finalize~&\Big[~\alpha~\Big]
	\end{aligned}
\end{gather}

The \finalize transaction only requires the parameter $\alpha$, identifying the PoI, because the blockchain already contains all necessary information about the PoI. The precondition of $t_1$ being expired ($t > t_1$) is necessary for the \finalize transaction to avoid premature finalization.

The effect of the \finalize transaction is that the contest for the PoI referred to by its signature $\alpha$ is concluded. This means that the winning witness is awarded the witness reward, which, according to Section~\ref{sec:approach}, is 1~PBT in our current approach. Furthermore, the conclusion of the contest performs the actual transfer of PBT, i.e., $x$ PBT are deducted from the balance of $\wallet_s$, and $\wallet_d$ receives $x-1$~PBT ($x$ reduced by the witness reward). This action is executed on all blockchains, since \finalize is posted on all blockchains.

We define the winning witness to be the contestant with the lowest signature $\omega$ (i.e., with its value closest to zero). Since this signature cannot be influenced by the contestants~(because it is only formed from the PoI data and the contestants' private key), they have no way of increasing their chances of winning a particular contest, except for creating a large number of wallets (private keys). Such ``mining for wallets'' is not a violation of our protocol and no threat to its fairness, since doing so is computationally expensive, and therefore creates cost on its own. There exists a break-even point of the witness reward and the cost created by the creation of a large number of wallets~\cite{tast-wp3}. Effectively, this challenge is comparable to mining in Proof of Work (PoW) in that resources, i.e., computing power, can be traded for rewards.

In our example above, the witness with the lowest $\omega$ is $\wallet_w$, with $\omega = \texttt{0xC1}$. Therefore, this witness is awarded the witness reward. The final blockchain state is shown in Table~\ref{tab:final}. The balances of the competing contestants $\wallet_u$ and $\wallet_v$ remain zero. The expired PoIs are no longer shown for brevity.

\begin{table}
\caption{Final State After Witness Contest at $t > 61$}
\label{tab:final}
\chains{16mm}{
	\bbb{\va{60}}{\va{19}}{\vaa{1}}%
}{
	\bbb{\va{60}}{\va{19}}{\vaa{1}}%
}{
	\bbb{\va{60}}{\va{19}}{\vaa{1}}%
}
\end{table}

\subsection{Prevention of Double Spending}

A malicious sender might sign two different PoIs conflicting with each other. For instance, a sender owning $10$~PBT might create two PoIs, transferring $8$~PBT each, to two different wallets. Executing these transfers would reduce the sender's balance by $16$~PBT in total, resulting in $-6$~PBT.

In order to prevent such behavior, we introduce the \veto transaction. The \veto transaction can be called by any party noticing two conflicting PoIs (i.e., two PoIs with the same source, different destinations, and overlapping validity periods). Since such PoIs are forbidden by definition, the \veto transaction is used to penalize the sender, and to protect the receiver from losing PBT due to inconsistent balances.

Since the \veto transaction requires incentive, we propose to use the same technique as presented above, i.e., a contest. Any observer of a PoI conflict can report this conflict using the \veto transaction, and after the expiration of the veto validity period, the observer with the lowest $\omega$ signature is assigned a reward.

We therefore define the \veto transaction as shown in~(\ref{eq:veto}), where $\alpha$ refers to the original PoI, which is known to the blockchain because it has already been posted on a given blockchain, and the remaining data $\xfer{s}{d'}{x'}$ and $t_0', t_1', \alpha'$ describe the new, conflicting PoI. 

\begin{gather} \label{eq:veto}
	\begin{aligned}
	\wallet_w:\veto~&\Big[~\alpha,\,\xfer{s}{d'}{x'},\,t_0',\,t_1',\,\alpha'~\Big]_{\omega}
	\end{aligned}
\end{gather}

The \veto transaction, similar to \contest, is posted on all participating blockchains. Note that multiple observers can be expected to concurrently post \veto transactions. Therefore, it is possible that on one blockchain, a given PoI~(e.g., where $\alpha = \texttt{0x10}$) is posted first, and a second PoI~(e.g., where $\alpha' = \texttt{0x20}$) is presented as ``conflicting'' by a \veto transaction, while on another blockchain, the PoI where $\alpha = \texttt{0x20}$ is posted first, and the PoI with $\alpha' = \texttt{0x10}$ is posted in the \veto transaction as ``conflicting''. However, in the following, we define a behavior for the \veto transaction that still maintains consistency, regardless of the order of PoIs.

The preconditions for \veto are that $\alpha$ refers to a PoI already known to the blockchain, that the conflicting PoI is valid, and that the two PoIs are actually conflicting.

The effects of \veto are as follows: (i)~The sender of the conflicting PoIs loses all PBT, i.e., the balance is set to zero to penalize such protocol-violating behavior. (ii)~Any PoI which has a non-expired validity period (i.e., every PoI where $t < t_1$) is canceled. This means that no $\finalize$ transaction will be permitted for this PoI, the transfer itself will therefore not be executed, and no witness reward will be assigned. Finally, (iii)~a new contest is started, called the \emph{veto contest}. The veto contest is similar to a regular witness contest in that its purpose is the propagation of information (in this case, the information of conflicting PoIs).

In the following, we propose a possible implementation of such veto contest, however, its details (i.e., the definition of its validity period or the reward) are specifics which may be implemented differently.

We propose to use the same reward for the veto contest as for the regular witness contest (in our case, $1$~PBT). Since all PBT held by the sender are destroyed, and only $1$~PBT is assigned to the winner of the veto contest, all remaining PBT are lost. Furthermore, we propose the validity period expiration of the veto contest, $t_\veto$, to be defined as shown in~(\ref{eq:veto-t1}).

\begin{equation}
	t_\veto = \max(t_1, t_1') + \max(t_1-t_0, t_1'-t_0') \label{eq:veto-t1}
\end{equation}

The definition shown in~(\ref{eq:veto-t1}) states that the veto contest is valid until a point in time which is found by taking the later expiration time of the conflicting PoIs~($\max(t_1, t_1')$) and adding the longer validity period~($\max(t_1-t_0, t_1'-t_0')$). This is done to ensure that sufficient time is available for the veto contest. Again, we note that this is an implementation detail and other approaches (e.g., a fixed period) are also possible.

The veto contest is concluded by a \finveto transaction, defined as shown in~(\ref{eq:finveto}).

\begin{gather} \label{eq:finveto}
	\begin{aligned}
	\finveto~&\Big[~\alpha,\,\alpha'~\Big]
	\end{aligned}
\end{gather}

The effect of the \finveto transaction is similar to that of the \finalize transaction, except that no actual transfer is executed. The witness reward is again assigned to the veto contestant---that is, a wallet posting a \veto transaction---with the lowest $\omega$ signature in the \veto transaction. Similar to the \finalize, the \finveto transaction can be called by anyone, in particular, the winning veto contestant has monetary incentive in doing so.

%% file: sections/40_eval.tex
\section{Evaluation}
\label{sec:eval}
The approach presented in Section~\ref{sec:approach} introduces transactions which change the state of different blockchains within a blockchain ecosystem, according to given rules. This can be implemented using smart contracts, e.g., using the Solidity language~\cite{dannen2017introducing}---more specifically, the EVM---on the Ethereum blockchain. We use Solidity to create a reference implementation of the proposed protocol for evaluation\footnote{\url{https://github.com/pantos-io/dextt-prototype}}. However, other ways of implementing such transactions exist. For instance, instead of using smart contracts (e.g., when dealing with blockchains without such capabilities), one might add backwards-compatible layers on top of blockchains, providing the required capabilities for the transactions presented in this work. A similar approach is used by OmniLayer~\cite{omnilayer} or CounterParty~\cite{counterparty,counterparty-protocol}, which add such layers for enhanced features. For this work, however, we use our reference Solidity implementation of DeXTT for evaluation and cost analysis, postponing the integration of approaches such as OmniLayer or CounterParty to future work. Nevertheless, our current evaluation is sufficient to demonstrate the overall functionality of the DeXTT protocol using Solidity smart contracts and the conceptual applicability.

In order to evaluate our approach, we investigate its functionality, performance, and cost impact in an ecosystem of blockchains with agents performing repeated token transfers. We achieve these goals by using our reference implementation consisting of Solidity smart contracts, deploying these smart contracts on a number of private Ethereum-based blockchains, and using testing client software to perform transfers with a given rate.

We ensure a reproducible and uniform ecosystem of blockchains by using three \texttt{geth} nodes in Proof of Authority~(PoA) mode, creating three private blockchains. We choose PoA to achieve an energy-efficient testing and evaluation platform while being able to perform repeated experiments. Note that the consensus algorithm, i.e., PoW, Proof of Stake~(PoS), or PoA, defines the behavior of blockchain nodes between each other and maintains data consistency in the network of a given blockchain~\cite{zheng2017overview}. However, the smart contract layer is independent of the consensus algorithm. Therefore, our evaluation on PoA is directly applicable to blockchains with any consensus algorithm, including PoW.

The \texttt{geth} nodes used in our experiments can be configured, for instance, with regards to block time and Gas limit. We observe the behavior of the live Ethereum blockchain (January 2019) and configure our nodes to follow this behavior. Therefore, our nodes are configured to use a block time of 13~s on average, and a Gas limit of 8~million Ethereum Gas, mimicking the live Ethereum chain. We use private chains instead of the Ethereum main chain to enable a high number and low cost of repeatable experiments in an automated fashion without depending on external components, such as Ethereum nodes.

We use 10 clients constantly and simultaneously initiating transfers within the blockchain ecosystem. This number is chosen as a balance between feasible and reproducible experiments and expected real-world conditions. While it is small compared to evaluations of other classes of distributed systems, we note that the lack of scalability of blockchain technologies is a crucial issue in general, and is seen as one of the main challenges for existing blockchain technologies~\cite{10.1007/978-3-319-45656-0_3}. We refer to existing literature for a study on how scalability of blockchains can be improved~\cite{vukolic2015quest}.

In our experimental ecosystem, each client constantly transfers random amounts of PBT to random wallets. If a client owns too little PBT for a transaction, no transaction is performed until PBT are available again. After a successful transfer, the client waits for a random time between 15~s and 30~s. Afterwards, the process is repeated indefinitely throughout the entire experiment duration.

We perform two experiment series, as described in the following sections. The first series is used to evaluate DeXTT scalability and the impact of the transfer validity period, and consists of a series of 30-minute experiments, where each individual experiment uses an increased validity period. The second series consists of 20 experiments, again with a duration of 30~min each, used to measure the average cost of a DeXTT transfer.

\subsection{Scalability and Timing}
\label{sec:scalability}

The DeXTT protocol requires one \claim transaction per transfer, and for each transfer, one \finalize transaction per blockchain. In addition, each contestant posts one \contest transaction to each blockchain. We assume that candidates which no longer have a chance to win the witness contest~(because a candidate with a lower signature $\omega$ for the given transaction has already posted a \contest transaction) do not post \contest transactions to avoid cost. Assuming uniform distribution of $\omega$ values, as defined in Section~\ref{sec:hashes}, on the average case, each \contest transaction halves the space of remaining possible winning signatures $\omega$ (because the expected value of the uniform distribution is the arithmetic mean of the domain). Therefore, with each \contest transaction, the likelihood of another candidate existing with a lower $\omega$ is halved. Following from this, on average, $\log_2 n$ candidates will post a \contest transaction, where $n$ is the number of total observers.

Transfer time in the DeXTT protocol is directly impacted by the transfer validity period $[t_0, t_1]$ chosen by the sender. We therefore first evaluate the impact of the validity period. Using too short validity periods leads to corrupted transfers, i.e., transfers which cause permanently inconsistent balances, since observers cannot post \contest transactions in time. In such scenarios, eventual consistency between blockchains is not guaranteed. As stated above, we use a block time of 13~s, therefore, we start our experiments with 10~s, and increase the period by 5~s with each experiment. We then run our blockchain ecosystem for 30~min using each validity period and record the number of corrupted transactions. Note that we have to reset the inconsistent balances for wallets participating in a corrupted transaction in order to be able to run the experiments for 30~min.

\begin{figure}
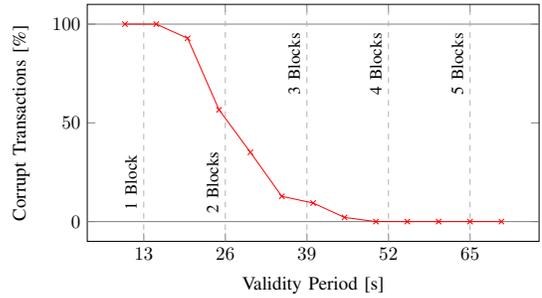

	\centering
	\includestandalone[width=0.8\columnwidth]{tikz/validity}
	\caption{Impact of Validity Period on Transaction Success}
	\label{fig:validity}
\end{figure}

Figure~\ref{fig:validity} shows the results of these experiments. Beyond 52~s, no corrupt transactions are observed. It becomes clear that using the reference implementation and waiting for 4 blocks~(52~s) is sufficient for ensuring consistency. Between 1 and 3 blocks~(13~s and 39~s, respectively), the amount of corrupted transactions declines with a varying rate.

From this experiment, we conclude that using a validity period with the length of at least 4 blocks~(52~s) is sufficient to maintain consistency using our reference implementation. Additional time may be required in order to accommodate slow network connectivity.

\subsection{Cost Analysis of DeXTT Transfers}
\label{sub:costanalysis}

To estimate the cost incurred by DeXTT transfers, we run the same experiment 20 times. Based on our previous experiment, we choose 65~s (5 blocks, well above the determined limit of 52~s) as the duration of the PoI validity period in each transaction. We record the average cost of each transaction. Table~\ref{tab:cost} shows an overview of the cost of the individual transactions involved in a DeXTT transfer. For each transaction, we show the mean cost, and its standard deviation, both in thousands of Ethereum Gas (kGas), and in USD. For this, we assume a Gas price of 10~Gwei (1 Ether = $10^9$~Gwei = $10^{18}$~wei) and a price of Ether of $115.71$~USD. These values were obtained from the Ethereum live chain in January 2019. Note that our implementation is optimized in that \claim and \contest both use the same smart contract function.
Nevertheless, we distinguish the semantic difference~(posting of new transfer for \claim, and participating in a contest for \contest) in the results.

\begin{table}
	\centering
	\caption{Cost Analysis}
	\label{tab:cost}
	\begin{tabular}{l r r c r r}
		\toprule
		  & \multicolumn{2}{c}{Cost (kGas)} && \multicolumn{2}{c}{Cost (USD)} \\
		\cmidrule{2-3} \cmidrule{5-6}
		Transaction	& Mean & $\sigma$ && Mean & $\sigma$\\
		\midrule
		\claim		&  $57.7$ & $11.1$ && $0.0668$ & $0.0128$ \\
		\contest	&  $81.5$ & $64.2$ && $0.0943$ & $0.0743$ \\
		\finalize	&  $45.5$ &  $0.1$ && $0.0527$ & $<0.0001$ \\
		\veto		& $131.3$ & $91.9$ && $0.1520$ & $0.1063$ \\
		\finveto	&  $48.6$ &  $1.7$ && $0.0563$ & $0.0020$ \\
		\bottomrule
	\end{tabular}
\end{table}

In the following, we assume $m$ blockchains and $n$ total observers. For our calculation, we assume that all observers monitor all blockchains, and post \contest transactions if it benefits them. A regular DeXTT transfer (i.e., one which does not contain a conflicting PoI, and therefore requires no veto) consists of one \claim transaction (on the target chain), $\log_2 n$ \contest transactions (as discussed in Section~\ref{sec:scalability}) on each blockchain, i.e., $m \log_2 n$ \contest transactions, and $m$~\finalize transactions. The \claim transaction is posted by the receiver, and each \contest transactions is posted by an observer (thus becoming a contestant). While the \finalize transaction can be posted by any party, posting it is beneficial to the receiver~(because it finalizes the transfer to the receiver), and therefore it can be expected that the receiver will bear its cost to finalize the transfer.

The expected cost in kGas for a DeXTT transfer are as follows: The receiver bears the cost for one \claim transaction~($57.7$~kGas) and $m$ \finalize transactions ($45.5$~kGas each). Each of the $\log_2 n$ expected observers posting transactions bears the cost for $m$ \contest transactions ($81.5$~kGas each). The sender does not bear any cost.

Assuming a blockchain ecosystem of 10~blockchains, the total transaction cost for the receiver is $0.59$~USD. Each of the $\log_2 n$ observers posting transactions bears cost of $0.94$~USD. These numbers represent our current reference implementation and can be regarded as an upper bound for DeXTT transfer cost. Any additional optimization to the smart contract code has the potential to further reduce the Gas cost of the individual transactions, and therefore, of the overall DeXTT transfer.

Additionally, these numbers allow us to reason about the economic impact of a currency using DeXTT transactions. Observers pay transaction cost of $0.94$~USD, and potentially receive a witness reward, currently defined as 1~PBT. The chance of an observer winning is $\frac{1}{n}$, however, according to the discussion in Section~\ref{sec:scalability} on average, only $\log_2 n$ out of all $n$ observers are expected to post \contest transactions. Therefore, the likelihood for an observer posting a transaction to win the contest is $\frac{\log_2 n}{n}$.

Therefore, the investment for each observer is $0.94$~USD, the contest reward is 1~PBT, and the winning likelihood is $\frac{\log_2 n}{n}$. From this, it follows that in order for the observer to have incentive to post \contest transactions in an ecosystem of $m = 10$ blockchains, the inequation shown in~(\ref{eq:ineq-p}) must hold, where $p$ is the price of PBT in USD.

\begin{equation}
\frac{\log_2 n}{n}\,p > 0.94\ \text{[USD]} \label{eq:ineq-p}
\end{equation}

In other words, the price of PBT in USD divided by the number of observers must be higher than $0.94$. Assuming $n = 10$ observers, the PBT price must be above $2.83$~USD. Assuming $n = 100$, the PBT price must be above $14.15$~USD. For $n = 1000$, the PBT price must be above $94.32$~USD. 

Note that these number assume $m = 10$ blockchains, and a fixed reward of 1~PBT. A pro rata reward, e.g., $1\%$ of the transferred PBT, would reduce the required PBT price but increase the complexity of calculating the witness incentive. Furthermore, a dynamic reward adaption based on the number of observers, similar to the variable mining rewards in Bitcoin, or a value selected by the sender, similar to the Gas price in Ethereum, can also be used to reduce the required PBT price, and therefore incentivize observers. We note again that these numbers pose an upper boundary for the expected DeXTT transfer cost and PBT price requirements for witness incentive.

%% file: tikz/validity.tex
\begin{tikzpicture}
	\begin{axis}[
		width=10cm,height=6cm,
		xlabel={Validity Period [s]},
		ylabel={Corrupt Transactions [\%]},
		xtick={13,26,39,52,65}]

	\draw[ultra thin,black!50!white] (axis cs:\pgfkeysvalueof{/pgfplots/xmin},0) -- (axis cs:\pgfkeysvalueof{/pgfplots/xmax},0);
	\draw[ultra thin,black!50!white] (axis cs:\pgfkeysvalueof{/pgfplots/xmin},100) -- (axis cs:\pgfkeysvalueof{/pgfplots/xmax},100);

	\draw[dashed,black!30!white] (axis cs:13,\pgfkeysvalueof{/pgfplots/ymin}) -- (axis cs:13,\pgfkeysvalueof{/pgfplots/ymax});
	\draw[dashed,black!30!white] (axis cs:26,\pgfkeysvalueof{/pgfplots/ymin}) -- (axis cs:26,\pgfkeysvalueof{/pgfplots/ymax});
	\draw[dashed,black!30!white] (axis cs:39,\pgfkeysvalueof{/pgfplots/ymin}) -- (axis cs:39,\pgfkeysvalueof{/pgfplots/ymax});
	\draw[dashed,black!30!white] (axis cs:52,\pgfkeysvalueof{/pgfplots/ymin}) -- (axis cs:52,\pgfkeysvalueof{/pgfplots/ymax});
	\draw[dashed,black!30!white] (axis cs:65,\pgfkeysvalueof{/pgfplots/ymin}) -- (axis cs:65,\pgfkeysvalueof{/pgfplots/ymax});

	\node[anchor=south,rotate=90] at (axis cs:13,20) {\small 1 Block};
	\node[anchor=south,rotate=90] at (axis cs:26,20) {\small 2 Blocks};
	\node[anchor=south,rotate=90] at (axis cs:39,80) {\small 3 Blocks};
	\node[anchor=south,rotate=90] at (axis cs:52,80) {\small 4 Blocks};
	\node[anchor=south,rotate=90] at (axis cs:65,80) {\small 5 Blocks};

	\addplot[color=red,mark=x] coordinates {
		(10,100)
		(15,100)
		(20,92.8)
		(25,56.6)
		(30,35.1)
		(35,12.8)
		(40,9.4)
		(45,2.1)
		(50,0)
		(55,0)
		(60,0)
		(65,0)
		(70,0)
	};
	\end{axis}
\end{tikzpicture}

%% file: sections/50_rw.tex
\section{Related Work}
\label{sec:rw}

As discussed in Section~\ref{sec:intro}, cross-blockchain interoperability can be used to address the fragmentation of the blockchain research field. Yet, to the best of our knowledge, contemporary approaches provide only limited interoperability across blockchains.

Initial interoperability was limited to trading assets on centralized exchanges. Subsequently, decentralized exchanges such as Bisq~\cite{bisq} or 0x~\cite{0x} emerged. Most recently, the Republic protocol~\cite{republic} has been proposed, which includes a decentralized dark pool exchange, i.e., details about an exchange are kept secret.

All of these approaches, however, are concerned with the \emph{exchange} of assets, generally using atomic swaps~\cite{herlihy2018atomic} for trustless asset exchange. In such an atomic swap, one party might transfer, e.g., Bitcoin to another party, while the other party transfers, e.g., Ether to the first. In such an atomic swap, each asset remains on its blockchain. In contrast, we propose a protocol for trading assets independently of a specific blockchains. In our approach, the balance information for such assets is stored on all blockchains simultaneously.

Another approach for a multi-blockchain framework is presented in PolkaDot~\cite{polkadot}. PolkaDot aims to provide ``the bedrock relay-chain'' upon which data structures can be hosted. However, in contrast to our approach, no specifics about cross-blockchain asset transfers are provided. Instead, PolkaDot is explicitly not designed to be used as a currency~\cite{polkadot}. Furthermore, PolkaDot is meant to be used as a basis for future blockchains (and other decentralized data structures), while in our current approach, we aim to use existing blockchains and implement functionality on top of them. However, the concepts presented in the PolkaDot paper are complementary to techniques we used in our approach.

Decentralized cross-blockchain transfers allow users to fully utilize the existing variety of blockchains, instead of being locked to a single blockchain. To the best of our knowledge, the approach closest to the work at hand is Metronome~\cite{mtn}, which uses assets available on multiple blockchains. However, Metronome proposes that assets still lie on one specific blockchain at a time, while in our proposal, the assets are not bound to one blockchain.

The DeXTT protocol presented in this paper is based on our own former work. The XPP has been formally described in~\cite{tast-wp2}. Furthermore, in~\cite{tast-wp3}, we describe the deterministic witness selection approach conceptually. The work at hand significantly extends our former work by providing a concrete implementation of this approach within the DeXTT protocol. Furthermore, the work at hand is a significant enhancement of our earlier work, in which we also defined a token existing across blockchains. However, each wallet had a different balance on each blockchain (and all balances were recorded on all blockchains) \cite{tast-wp2}. In the work at hand, this concept is simplified, yielding only one balance per wallet~(which is recorded on all blockchains).

%% file: sections/60_conclusion.tex
\section{Conclusion}
\label{sec:conclusion}

In this paper, we have presented DeXTT, a protocol for transferring cross-blockchain tokens, existing on a single blockchain, but tradeable on multiple blockchains. This reduces dependency on a single blockchain and risk, e.g., of selecting a blockchain which later suffers from a security breach. DeXTT ensures eventual consistency of balances across blockchains, and prohibits double spending. We have presented the protocol in detail, implemented it in Solidity, and provided an experimental evaluation, highlighting its performance with regards to time and cost.

Our evaluation shows that the reference implementation of DeXTT requires at least 4 blocks for maintaining consistent balances. Furthermore, we show that a DeXTT transfer using our reference implementation costs 103.2~kGas for the receiver, and 81.5~kGas for any contributing observer. We also provide an analysis of the economic impact of the witness rewards based on the parameters of the multi-blockchain ecosystem used.

In future work, we will address the main limitation of our current evaluation by implementing DeXTT using additional technologies such as OmniLayer or CounterParty and therefore evaluate the performance of DeXTT in a blockchain ecosystem consisting of mixed blockchain types. Furthermore, we aim to implement DeXTT on other native smart contract platforms such as EOS.IO~\cite{eosv2}. In addition, we aim to evaluate more refined approaches for the veto contest, which can be used to relax the currently strict requirement of not signing to outgoing PoIs with overlapping validity periods.